\documentclass[aps,prb,twocolumn,superscriptaddress,showpacs]{revtex4}

\usepackage{graphicx}
\usepackage{dcolumn}
\usepackage{bm}

\begin{document}
\title{Magnetoelectric coupling tuned by competing anisotropies in Mn$_{1-x}$Ni$_{x}$TiO$_3$}

\author{Songxue Chi}
\affiliation{Quantum Condensed Matter Division, Oak Ridge National Laboratory,
Oak Ridge, Tennessee 37831, USA}
\author{Feng Ye}
\affiliation{Quantum Condensed Matter Division, Oak Ridge National Laboratory,
Oak Ridge, Tennessee 37831, USA}
\affiliation{Department of Physics and Astronomy,~University of Kentucky, Lexington, Kentucky 40506, USA}
\author{H.~D.~Zhou}
\affiliation{Department of Physics and Astronomy, University of Tennessee, Knoxville, Tennessee 37996-1200, USA}
\affiliation{National High Magnetic Field Laboratory, Florida State University, Tallahassee, FL 32310-3706, USA}
\author{E.~S.~Choi}
\author{J. Hwang}
\affiliation{National High Magnetic Field Laboratory, Florida State University, Tallahassee, FL 32310-3706, USA}
\author{Huibo Cao}
\affiliation{Quantum Condensed Matter Division, Oak Ridge National Laboratory,
Oak Ridge, Tennessee 37831, USA}
\author{Jaime~A.~Fernandez-Baca}`
\affiliation{Quantum Condensed Matter Division, Oak Ridge National Laboratory,
Oak Ridge, Tennessee 37831, USA}
\affiliation{Department of Physics and Astronomy, University of Tennessee, Knoxville, Tennessee 37996-1200, USA}
\date{\today}

\begin{abstract}

A flop of electric polarization from $P$$\parallel$$c$ ($P_c$) to $P$$\parallel$$a$ ($P_a$) is observed in MnTiO$_3$
as a spin flop transition is triggered by a $c$-axis magnetic field, $H_{\|c}$=7 T. 
The critical magnetic field $H_{\|c}$ for $P_a$ is significantly reduced in Mn$_{1-x}$Ni$_x$TiO$_3$ (x=0.33).  
$P_a$ and $P_c$ have been observed with both 
$H_{\|c}$ and $H_{\|a}$. Neutron diffraction measurements revealed
similar magnetic arrangements for the two compositions where
the ordered spins couple antiferromagnetically with their nearest 
intra- and inter-planar neighbors.
In the x=0.33 system, the uniaxial and planar anisotropies of Mn$^{2+}$  and Ni$^{2+}$
compete and give rise to a spin reorientation transition at $T_R$.
A magnetic field, $H_{\|c}$, aligns the spins along $c$ for $T_R$$<$$T$$<$$T_N$. 
The rotation of the collinear spins away from the $c$-axis for $T$$<$$T_R$ 
alters the magnetic point symmetry and gives rise to a new ME susceptibility tensor form. 
Such linear ME response
provides satisfactory explanation for the behavior of the field-induced electric polarization in both compositions. As the Ni content
increases to  x=0.5 
and 0.68, the ME effect disappears as a new magnetic phase emerges. 

\end{abstract}

\maketitle

\begin{center}
${\bf I.~~INTRODUCTION}$
\end{center}

The multiferroics that show strong magnetoelectric (ME) effect are among
the most saught-after materials due to their multi-functionality of inducing
polarization with magnetic field or magnetization with electric field. \cite{Fiebig,Cheong,Eerenstein,Khomskii,Kleemann,Kimura3} 
The linear ME effect occurs in a crystal when the term $-\alpha_{ij}E_iH_j$ 
in the expansion of its Gibbs free energy $F(E,H)$ is nonzero. Here $\alpha$
is a second rank tensor which changes sign under space or time 
inversion, but is invariant when the occurrences of the two inversions 
are simultaneous. \cite{Agyei, Rivera}
The magnetic symmetries that meet such conditions are allowed to have 
linear ME response. Therefore in    
exploring magnetoelectrics among materials with long range magnetic order,
symmetry analysis serves as a reliable guide. \cite{Harris,Zvezdin,Bousquet1} Such predictability 
can be obscured when extra microscopic complications, such as
magnetic anisotropy, spin frustration and spin-lattice coupling, have been 
introduced. However, these extra variables 
sometimes help to increase the magnitude of $\alpha$ \cite{Mostovoy, Bousquet1, Bousquet2} 
or even give rise to new ME
coupling mechanisms. \cite{Hornreich, Wojdel, Yamauchi}

\begin{figure}[bt!]
\includegraphics[width=.94\columnwidth]{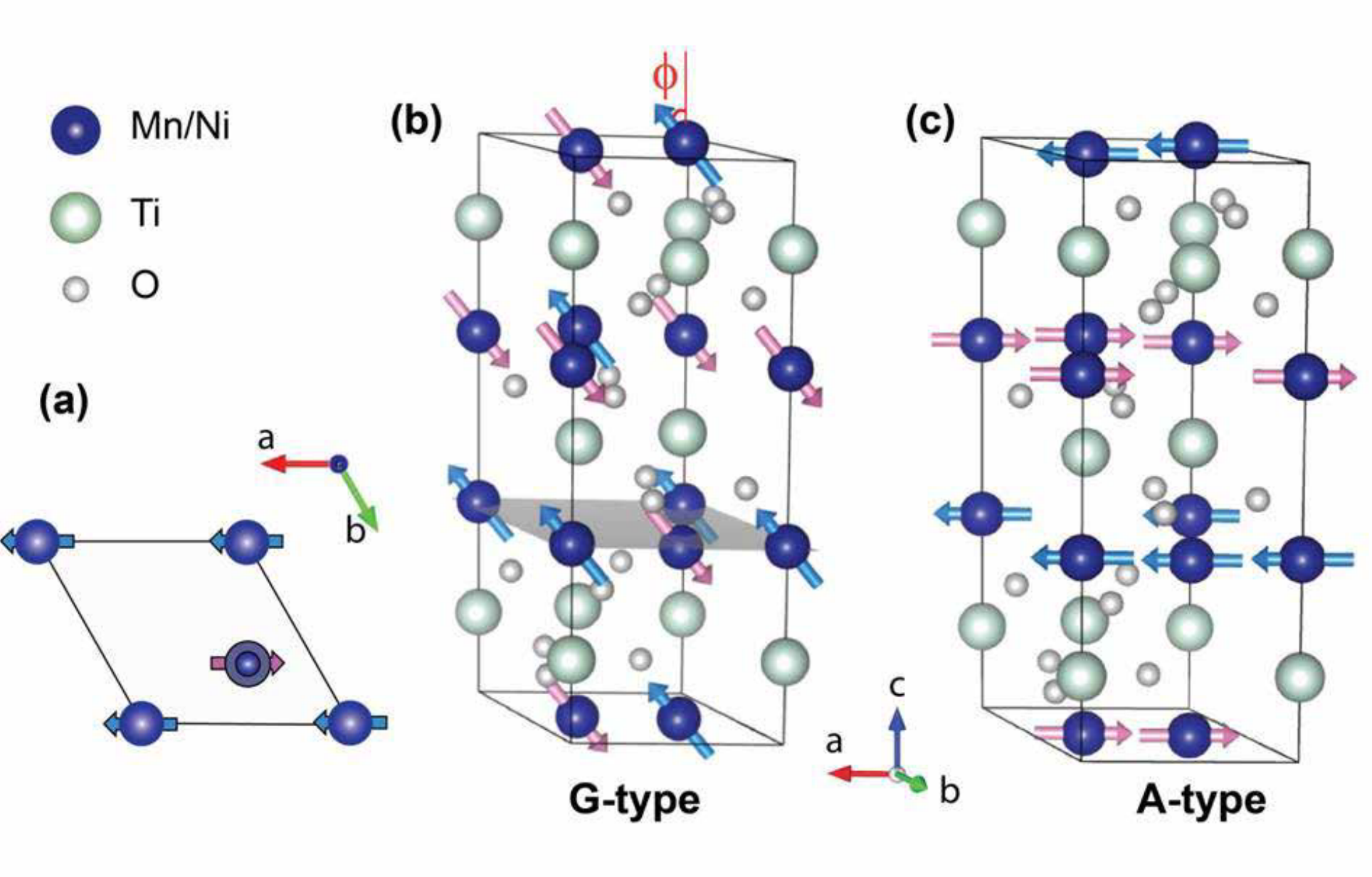}
\caption{(Color online) Primitive cells of the hexagonal lattice of the
compounds Mn$_{1-x}$Ni$_x$TiO$_3$. (a)  The top view of the hexagon of cations 
with respect to the rhombic cross section as shaded in (b).
Note the vacant octahedral site and a cation displacement from the ab-plane. 
Two types of magnetic structures, G-type with real spin directions 
unspecified and A-type, are shown in (b) for Mn-rich compounds and (c) for the Ni-rich compounds.  
$\phi$ in (b) is the angle between the spins and the hexagonal $c$-axis.
}
\label{fig1}
\end{figure}

Mn$_{1-x}$Ni$_x$TiO$_3$ 
is such a system where more than one ME mechanism has emerged.
MnTiO$_3$ and NiTiO$_3$ have the same ilmenite
structure (Fig. 1) \cite{Ishikawa} but different spin arrangements and easy axes, \cite{Shirane} 
which compete in the mixed compounds. Also competing
are the energy loss from single ion anisotropy and that from the 
frustration of the exchange interactions. \cite{Yoshizawa} 
Various new magnetic phases including spin glass (SG) phase emerge as 
a result, forming a rather complex
phase diagram. \cite{Ito_phaseDiagram, Yoshizawa, Kawano} 
While the linear ME effect was observed in MnTiO$_3$ as its 
magnetic symmetry permits, \cite{Mufti} a new ME response 
is induced by the toroidal moments in the SG state of the mixed compounds. 
 \cite{Yamaguchi_2012} On both sides of the SG phase, 
the unexplored spin flop transitions, short range magnetic correlations  
and the Ti$^{4+}$ ions with empty 3$d$ shells, \cite{Cohen,Deng} are all potential 
hosts of yet another novel ME mechanism.   
Although the magnetic structures of the end-member compounds have been studied, \cite{Shirane}
the details of the magnetic evolution in the mixed compounds and its effect on the electric   
polarization are still lacking.   
This report presents a systematic investigation of the ME effects and 
the magnetic orders in Mn$_{1-x}$Ni$_x$TiO$_3$. 
New components of the ME tensor and an anomaly 
in their temperature dependence under a low magnetic field       
have been observed. Neutron diffraction
measurement on the x=0.33 compound under an applied magnetic field reveals the effect of 
the field on the spin orientation and therefore, on the nature of the new ME coupling.    
Details of the magnetic orders in 4 typical compositions and corrections to 
the phase diagram are reported. 

\begin{center}
${\bf II. ~~EXPERIMENTAL}$
\end{center}

Single crystals of Mn$_{1-x}$Ni$_{x}$TiO$_3$ (x=0, 0.33, 0.50 and 0.68) 
were grown by the 
traveling-solvent floating zone technique. For electric polarization measurements, 
silver epoxy was pasted on the crystals cut into thin plates. 
The pyroelectric current was measured using a Keithley 6517A electrometer on 
warming after poling the crystal in an electric field of 800 kV/m while cooling 
down from above $T_N$ . The spontaneous polarization was obtained by 
integration of the pyroelectric current with respect to time.The single crystal 
neutron diffraction measurements 
were carried out at the High Flux Isotope Reactor of the Oak Ridge
National Laboratory. The HB-2C Wide Angle Neutron Diffractometer (WAND) with
wavelength of 1.482 $\text{\AA}$ was
used for reciprocal space and diffuse scattering surveys. The
collections of reflections for structural determination
were carried out at HB-3A 
four circle diffractometer where the wavelength of 1.542 $\text{\AA}$ was chosen.
 An assembly of permanent magnets  
that provides 0.7 Tesla at the sample position was employed in the magnetic field 
measurement on HB-3A. 
Closed-cycle refrigerators were used on both diffractometers.
The Rietveld refinements on the crystal and magnetic structures were conducted
using the FullProf Suite. \cite{FullProf} 

\begin{center}
{\bf III. ~~RESULTS}
\end{center}

\begin{center}
{\bf A. ~~Pyroelectric measurements under magnetic field}
\end{center}

\begin{figure}[bt!]
\includegraphics[width=0.94\columnwidth]{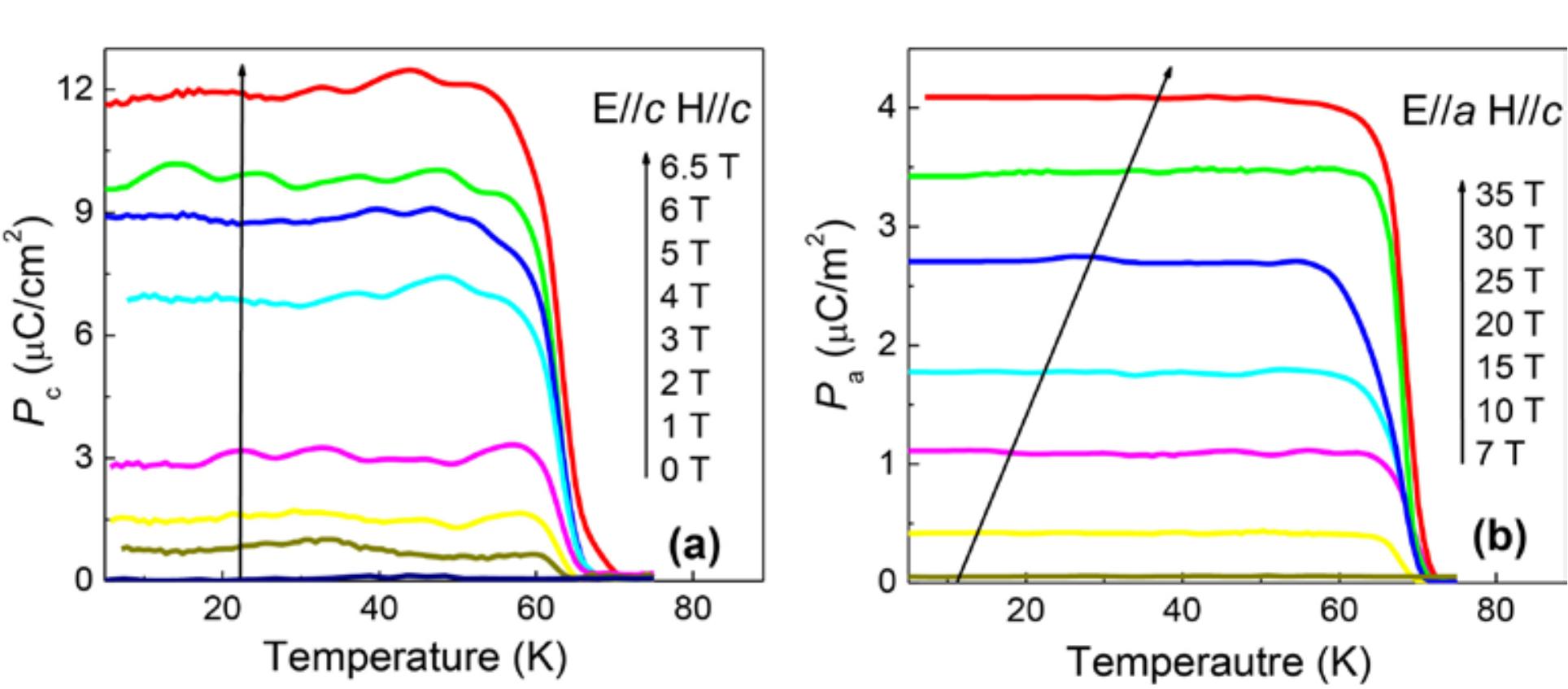}
\caption{(Color online) Temperature dependence of the electric polarization $P$ of MnTiO$_3$ at various magnetic fields measured with 
$H$$\parallel c$ and (a) $E$$\parallel c$, and (b) $E$$\parallel a$. The temperature dependence of $P$ for the x=0.33 compound with $H$$\| c$ and
(c) $E$$\parallel c$, and (d) $E$$\parallel a$. 
}
\label{fig2}
\end{figure}

\begin{figure}[bt!]
\includegraphics[width=.94\columnwidth]{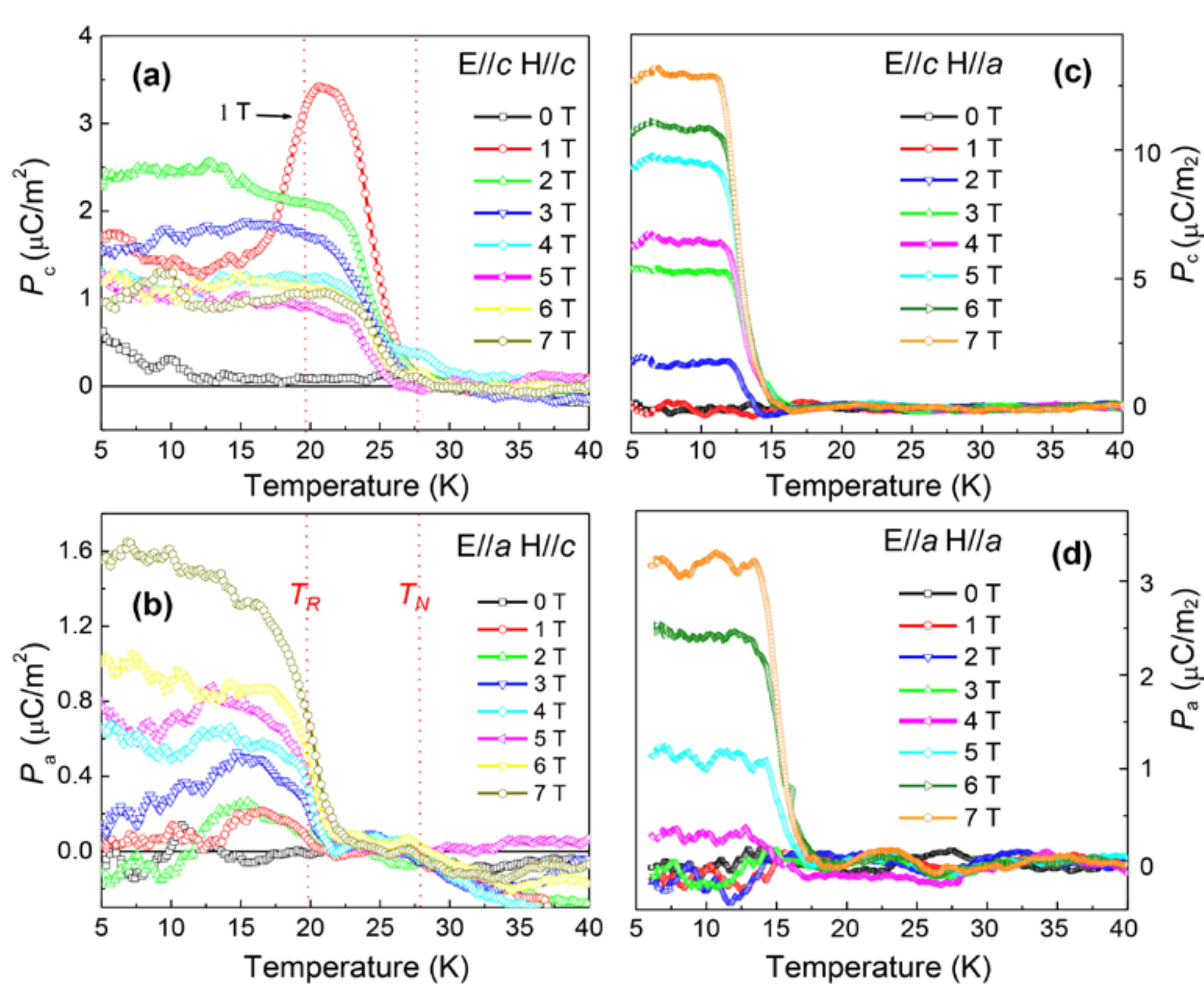}
\caption{(Color online) Temperature dependence of the electric polarization $P$ of Mn$_{0.67}$Ni$_{0.33}$TiO$_3$ at various magnetic fields measured with 
$H$$\parallel c$ and (a) $E$$\parallel c$, and (b) $E$$\parallel a$. The temperature dependence of $P$ for the x=0.33 compound with $H$$\parallel a$ and
(c) $E$$\parallel c$, and (d) $E$$\parallel a$. The dotted red lines mark the N$\acute{e}$el temperature and the onset temperature of the spin rotation
under the magnetic field of 0.7 T.
}
\label{fig3}
\end{figure}
The ME effect was observed in x=0 and 0.33.  
In both cases the pyroelectric current anomaly signaling the onset of polarization
appears only when finite magnetic field is applied along $c$. 
Fig. 2 shows the temperature dependence 
of the spontaneous polarization ($P$) under various magnetic fields for the x=0 sample. 
The previous study \cite{Mufti} only reported 
the observation of $P_c$ in this compound. As shown in Fig.2(a), $P_c$ increases monotonically with increasing field until $H_{\|c}$=6.5 T, then 
starts to decreases quickly and disappears above 7 T. The maximum value of $P_c$ is about 12 $\mu C/m$, which is consistent with
ref \cite{Mufti}. The intensity of $P_a$, on the other hand,
appears and starts to grow only above 7 T, as shown in 
Fig. 2(b). Magnetic field of 7 T along $c$ is where a spin flop in the magnetization was reported. \cite{Yamauchi} 
The magnetic field induced $P$ can be attributed to linear ME effect for several reasons: 
(1) Polarizations for both directions are linearly dependent on $H_{\|c}$. 
(2) The G-type magnetic structure with spins along $c$ belongs to $\bar{3}^{\prime}$ point group which 
does permit a non-zero $\alpha_{zz}$. (3) A dielectric anomaly appears in the vicinity of $T_N$. \cite{Mufti}
The switch of polarization from $P_c$ to $P_a$ signifies the change of the ME tensor, and therefore of the
magnetic symmetry. The ME coefficient $\alpha_{xz}$ and $\alpha_{zz}$, deduced from the slope of the $P$-$H$ curve, are 
4.44$\times 10^{-6}$ and 5.1$\times 10^{-5}$ (CGS unit), respectively. These values are about an 
order of magnitude smaller than those of Cr$_2$O$_3$. \cite{Popov, Rivera_2}

In the x=0.33 compound, the magnetic field induced polarization persists, but its behavior differs from that in undoped compound.
The threshold of field $H_{\|c}$ for $P_a$ 
disappears. Both $P_c$ and $P_a$ start to increase as soon as $H_{\|c}$ is turned on, as shown in Fig.3(a) and (b). 
$P_a$ linearly increases with $H_{\|c}$  (Fig.3(b)), but $P_c$ increases first then decreases to 1 $\mu C/m^2$ and remains
unchanged from 4 T to 7 T (Fig.3(a)). The polarizations along the two different $E$ directions also have different 
temperature dependence. The onset temperature of $P_a$ is at about 20 K while that of $P_c$ is 27 K. Moreover, at $H_{\|c}$ =1 T
the initial increase of $P_c$ on cooling is considerably suppressed below 20 K, as shown by the red circle in Fig.3(a). 
Such an anomaly is absent for higher fields. Additionally, the $P_a$ and $P_c$
can also be induced by $H$$//$$a$ (Fig. 3(c) and (d)), which does not give rise to any polarization in the x=0 compound.
Compared to the $H_{\|c}$-induced polarizations, the onset temperature for the palarizatoin with $a$-axis field is different though. 
In Fig.3(c) and (d), $P_c$ and $P_a$ both appear below 17.5 K. 
Different critical values of $H_{\|a}$  are required for $P_c$ and $P_a$, which are around 2 Tesla and 4 Tesla, respectively. Above the critical 
$H_{\|a}$ , the polarization increases with the $H_{\|a}$  in both cases. 
The polarization was not observed in the x=0.50 and 0.68 crystals regardless of the directions and magnitudes of the applied magnetic field. 
The knowledge of detailed spin structures in these mixed compounds and their evolution with temperature and magnetic field is 
needed to understand the coupling of the ferroelectric order with the magnetic one.

\begin{figure}[bt!]
\includegraphics[width=.94\columnwidth]{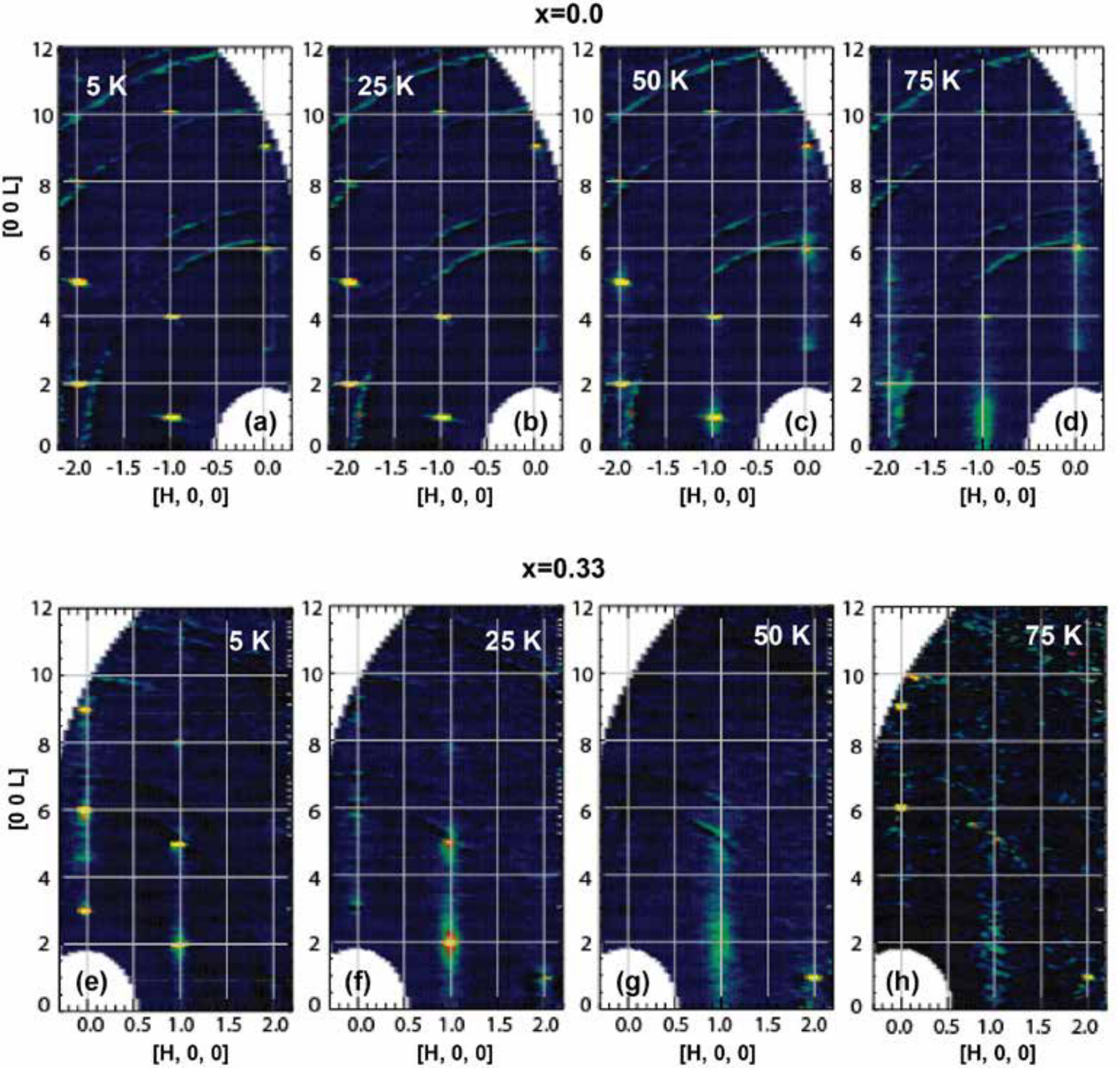}
\caption{(Color online) The contour plot of the diffraction pattern in the (H,0,L) plane
for x=0 at (a) 5 K, (b) 25 K, (c) 50 K, and (d) 75 K. (e)-(h) show the same plane for x=0.33 at 5 K, 25 K, 50 K and 75 K, respectively. 
The intensity scale in (h) is different from other panels to show the diminishing diffuse scattting. 
}
\label{fig4}
\end{figure}

\begin{center}
${\bf B. ~~The~Ni-doping~dependence~of~magnetic~order}$
\end{center}

\begin{center}
$\textbf{\textit{1.~~The~G-type~AFM~phase}}$
\end{center}

The structural refinements show that the 4 compositions of Mn$_{1-x}$Ni$_x$TiO$_3$ 
compounds all crystallize in space group
$R \bar{3}$. Their ilmenite structure and the two generalized
spin configurations are depicted in Fig.1. 
Along the $c$-axis of the hexagonal lattice, Mn$^{2+}$/Ni$^{2+}$ and 
Ti$^{4+}$ layers alternate and every third octahedral site 
is vacant. 
The magnetic structure of MnTiO$_3$ is  G-type
where nearest inter- and intra-planar neighbor spins are antiparallel, \cite{Shirane} 
which has the propagation wave vector $\vec{q}$=(0,0,0). The magnetic peaks coincide with the 
allowed nuclear ones (-$H$+$K$+$L$=$3n$, $n$ is integer). 
The contour plots of the magnetic diffraction  
in the ($H0L$) scattering plane, obtained by 
subtracting the high temperature (140 K) data 
as background, are shown in Fig. 4(a-d).
The temperature dependence of the (1,0,1) position 
gives the N$\acute{e}$el temperature $T_N$ $\sim$ 64$\pm$ 2.4 K.
The absence of peaks along [0,0,L] implies that the Mn$^{2+}$
moments are along $c$.   
The ridge-like diffuse scattering along $c$ starts to develop around 90 K.
Fig.4(d) show the diffuse peaks at 75 K, which center on 
the magnetic Bragg peak positions such as (1,0,1),
instead of (1,0,0). \cite{Akimitsu}  On cooling the diffuse scattering intensity
reaches its maximum at $T_N$, then quickly decreases. \cite{Akimitsu2}
Before Lorentzian peaks completely disappear at 4 K, they coexist 
with the Gaussian line shape,
suggesting the coexistence of long-range AFM order 
and short range 2D AFM correlations. 

\begin{figure}[bt!]
\includegraphics[width=1\columnwidth]{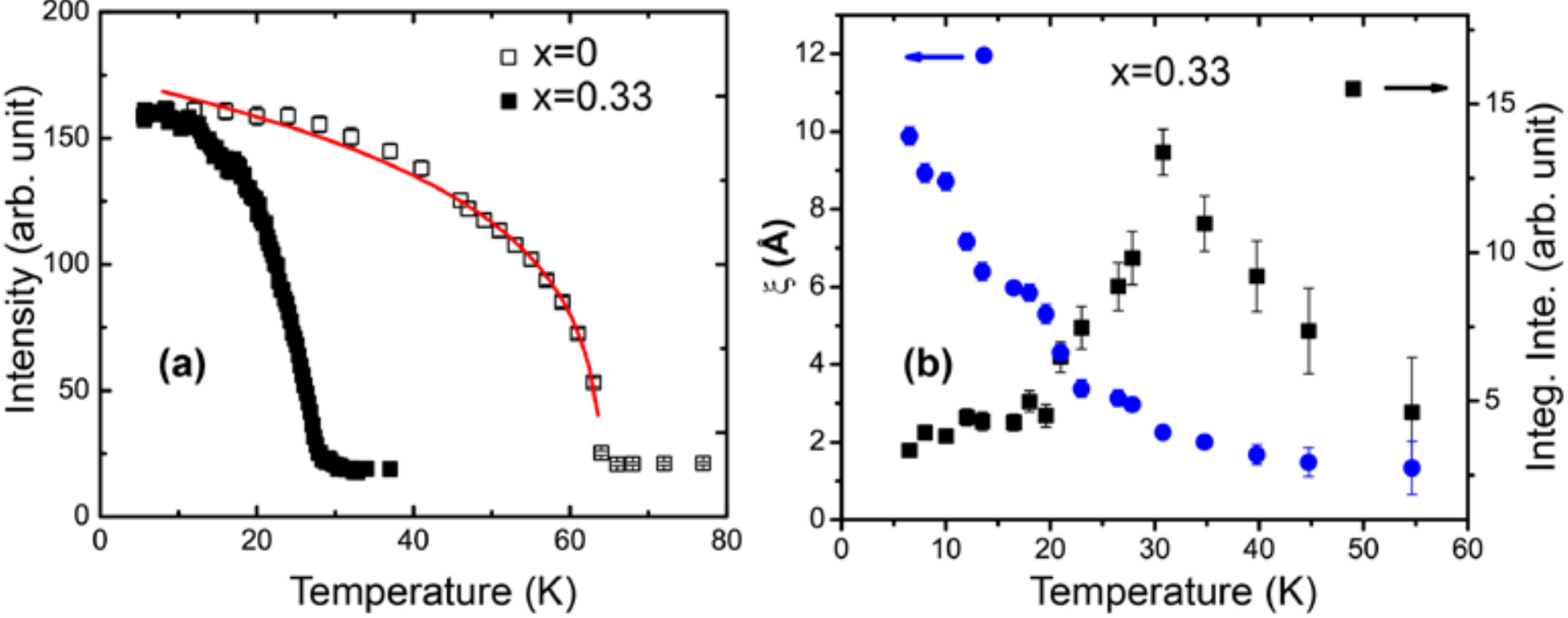}
\caption{(Color online)  (a) The intensity of the magnetic peak (1 0 1) as 
a function of temperature in x=0 and x=0.33. (b) The integrated
intensity of the diffuse scattering and short-range correlation length along $c$ as 
a function of temperature in x=0.33.  
}
\label{fig5}
\end{figure}

The spin structure of x=0.33 system remains G-type as suggested by the 
unchanged magnetic peak positions in Fig. 4(e). The onset temperature of the AFM order
is suppressed by Ni-doping to 27.6 K (Fig. 5(a) and Fig. 6(a)). 
However, the temperature dependence of the magnetic peaks, shown in Fig.5(a) and Fig.6(a),
indicates an extra phase transition at $T_R$=17.5 K. Both (0,1,2) and (1,0,1) show a kink at 
this temperature and (0,0,3) suddenly gains intensity below $T_R$ suggesting the spins rotate
away from the $c$-axis and obtain the component of the moment
perpendicular to the wavevector. To accurately characterize the magnetic configuration and monitor the
changing spin directions, 
116 magnetic Bragg peaks were collected for every 1 K between
5 K and $T_N$. In the magnetic structure refinement using FullProf, three equivalent magnetic domains 
were taken into account, only one of which is presented here.
The component of the ordered moment in the $ab$-plane at 
all measured temperatures lies in the $a$-direction. So the spin directions
are specified by $\phi$, the angle between the spin and the $c$-direction in the $ac$-plane,
as shown in Fig.1(b). The blue up-triangles in Fig.6(d) show the spin orientation $\phi$ as a function of temperature.
The ordered spins between $T_N$ and $T_R$ are close to but not quite along $c$ 
($\phi$=14.26$^{\circ}$ at 20.5 K). 
Cooling across $T_R$ the spins abruptly rotate by more than 60 degrees toward $a$.
The angle $\phi$ reaches 80.1$^{\circ}$ at 4 K. These results are different
from the established phase diagram which shows spins lying exactly along $a$ between
$T_N$ and $T_R$ and exactly along $c$ below $T_R$. \cite{Ito_phaseDiagram, Yoshizawa}
Fig. 6(c) shows the refined ordered moment as a function of temperature, which   
is a smooth decrease and proves that the kinks of the magnetic peak
intensities at 20 K in Fig. 6(a) are solely caused by the reorientation of the spins.  

The diffuse scattering at this composition becomes more prevalent: The ridge along $c$ 
persists to the lowest
measured temperature, extends high above $T_N$, and becomes broader than the undoped 
system (Fig.4(g)). 
The integrated intensity of the diffuse component around (1,0,1) also reaches its maximum 
at $T_N$ and decreases quickly on both sides (Fig.5(b)). In addition, the Lorentzian line 
width does decrease on cooling. The inter-plane spin correlation length $\xi$
is smaller than the nearest neighbour interlayer distance above $T_N$, implying the short range order
is basicaly 2-dimensionl (2D). The crossover from 2D to 3D occurs close to $T_N$ when the correlation length becomes bigger than 
the distance between neighbouring Mn/Ni layers.
$\xi$ does not diverge at $T_N$ but continues to increase on cooling to the base temperature. 
With some short-range correlated spins participating in the establishment of 3-dimensional long range 
order, some remain short-ranged at low temperature.    
 
\begin{figure}[bt!]
\includegraphics[width=.94\columnwidth]{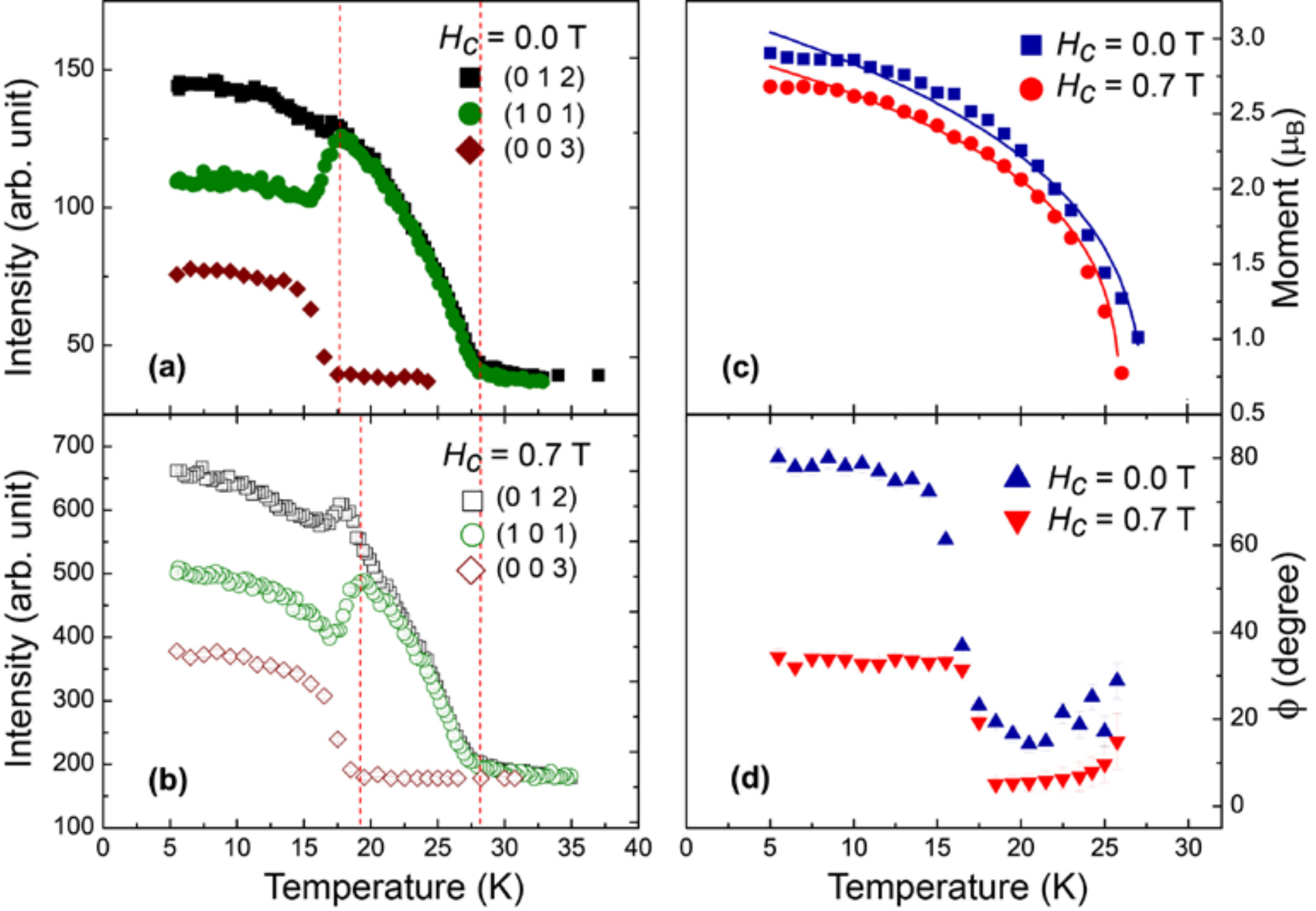}
\caption{(Color online) Temperature dependence of (a) various magnetic reflections of the x=0.33
compound without magnetic field, (b) magnetic peaks with applied external magnetic field of 0.7 T,
(c) the refined ordered moment and (d) the angle $\phi$ between the ordered spins and $c$. 
}
\label{fig6}
\end{figure}

\begin{center}
\textbf{\textit{2. Magnetic field effect on the AFM order (x=0.33)}}
\end{center}

The onsets of $P_a$ and $P_c$ occur at $T_R$ and $T_N$, respectively.
The anomalous suppression of $P_c$ under low field also coincides with $T_R$. 
Given that there is no detectable structural transitions at these temperatures, 
the electric polarization in the x=0.33 system apparently originates from the magnetic order. 
To investigate if this is linear or higher order ME effect, it is critical to know 
the effect of $c$-direction magnetic field on
the symmetry of the AFM order. The same crystal was aligned and mounted in the permanent magnet set which were mounted in
a CCR with the c-axis parallel to the field direction. The selected magnets provided a field of 0.7 T at the sample position, which was measured by Gauss meter.  The actual angle between 
$c$ and the field was determined to be 
6$^{\circ}$ using the observed angle $\chi$ of the Bragg peak (0,0,6). 
The temperature dependence of the magnetic peak intensities is shown in Fig.6(b).
Due to the geometrical restrictions imposed by the magnets, fewer magnetic peaks
were accessible, but enough were collected for an unambiguous refinement of the spin structure at each temperature. 
The field kept the spin structure and $T_N$
intact, but increased $T_R$ from 17.5 K to about 20 K, making it the same as the onset temperature for $P_a$. 
The result of spin structure refinements shows that the spins
were pulled toward $c$ by the field, both below and above $T_R$. $\phi$ is reduced to about 5$^{\circ}$
above $T_R$ and around 30$^{\circ}$ below. It is reasonable to assume that the spins would have been
aligned along the $c$-axis had a higher field been perfectly applied along $c$. 
The spin-rotation transition is made sharper by the small field. Another effect of 
this field is suppressing the moment as shown in Fig. 6(c). 

\begin{figure}[bt!]
\includegraphics[width=.94\columnwidth]{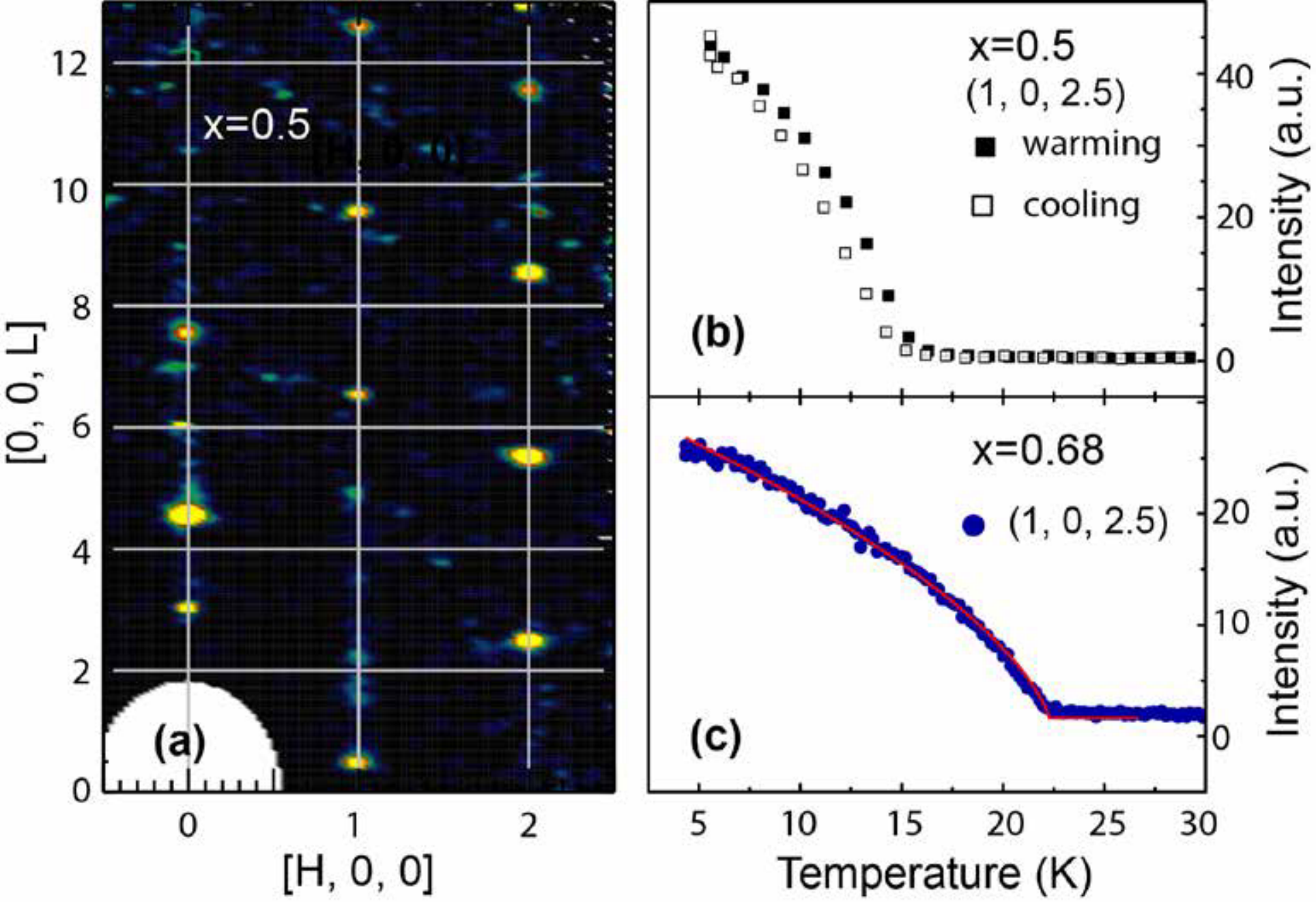}
\caption{(Color online) (a) Contour plot of the (H,0,L) scattering plane collected at 4 K with the 25 K data 
subtracted as the background. The visible (0,0,3) and (0,0,6) peaks do not show temperature dependence. 
The temperature dependence of the magnetic peak (1,0,2.5) for the (b) x=0.5 and (c) x=0.68 compound.
}
\label{fig7}
\end{figure}

\begin{center}
\textbf{\textit{3. A-type AFM structure (x=0.50 and  0.68)}}
\end{center}

The SG state forms between x=0.4 and 0.48 according to 
the established phase diagram.  \cite{Yoshizawa,Ito,Yamaguchi_2012}
The contour plot of the (H,0,L) scattering plane for x=0.5 
is taken at 4 K with 20 K data subtracted and shows a new magnetic wave vector $\vec{q}$=(0,0,1.5),
indicating the A-type magnetic order 
has taken over at this composition. There is no sign of diffuse scattering 
along $c$. The Bragg peak (1,0,2.5), shown in Fig.7(b), decreases smoothly in intensity
without an abrupt transition and completely vanishes above 15 K. The smeared transition also shows hysteresis on cooling
suggesting the spin glass phase still lingers at this composition. This is consistent with the magnetization measurement.
\cite{Ito} The coexistence of long range AFM order with the spin glass order has been predicted \cite {Sherrington} in such a 
magnetically non-diluted system. Similar phonomena have been observed in Mn$_{1-x}$Fe$_x$TiO$_3$, where the dominant nearest neighbor 
interactions compete with each other and give rise to a strong magnetic frustration within the honeycomb layer. \cite {Yoshizawa2, Katori}
The spin structure refinement 
agrees with the A-type model with
the spins lying along $a$-axis, as shown by Fig. 1(c).
As the nickel content increases to 0.68, the arrangement of the ordered moments remains A-type. Both the N$\acute{e}$el temperature
and the size of the ordered magnetic moment at low
temperature increase. $T_N$ increases to 21.5 K and the transition is abrupt 
and first-order like, contrasting with that of the x=0.5 sample. 
The structural parameters at 4 K and the magnetic orders in compounds of different Ni concentrations are 
summarized in Table 1. 

\begin{table}[ht!]
\caption{The lattice parameters, atom parameters, magnetic structures, magnetic phase transition temperatures, 
the ordered moments of the magnetic orders, and the R factors of the structure refinements in 
various Mn$_{1-x}$Ni$_x$TiO$_3$ compounds. m$_a$ and m$_c$ denote the projected moment on the hexagonal $a$- and $c$-axes respectively. 
$R_{F^2}$ is calculated by $R_{F^2}$=$100\sum_n(\left|G^{2}_{obs}-G^{2}_{calc}\right|)/\sum_nG^{2}_{obs}$, where $G$ is the 
structure factor and $n$ the number of reflections used.
}
\label{tab1}
\begin{ruledtabular}
\begin{tabular}{llccccc}
 refined x & x=0.00 \cite{Kidoh} & x=0.33  & x=0.50 & x=0.68 & x=1.00 \cite{Boysen} \\
		 \hline\hline
  $a$ & 5.14 & 5.12 & 5.06 &5.05&5.04  \\

  $c$ & 14.28 & 14.15 & 13.91 & 13.91 & 13.81  \\

  $Mn_z/Ni_z$    &  0.3600  &  0.347(2)  &  0.3471(2)  &  0.3509(5)  & 0.3509  \\

  $Ti_z$  & 0.1476 & 0.1504(5) & 0.1466(2) & 0.1426(8)& 0.1450 \\

  $O_x$ & 0.3189 & 0.3188(6) & 0.3166(2)  & 0.3161(8) & 0.3142  \\

  $O_y$ & 0.031 & 0.0264(8) & 0.0246(3) & 0.0205(7) & 0.016 \\

  $O_z$ & 0.2439 & 0.2449(3)  & 0.2459(1) & 0.2458(3) & 0.2465 \\
	
  magn.    &    &    &    &    &   \\
	struc.  & G-type & G-type & A-type & A-type & A-type \\

  $T_N$ & 64 & 27.6 & 15 & 21.5 & 21.8 \cite{Shirane} \\

  $T_R$ & $-$ & 17.51 & $-$ & $-$ & $-$ \\

  $m_a$ ($\mu_B$) & 0 &2.86(2)&1.36(3)& 2.06(3) & 2.25 \cite{Shirane}\\

	$m_c$ ($\mu_B$) & 4.55  \cite{Shirane} & 0.5(1) & 0 & 0 & 0\\

Nucl. $R_{F^2}$ &    &  7.77  &  6.33  & 8.05 &   \\

Magn. $R_{F^2}$ &    & 3.97   &  15.2  & 9.6 &   \\
  \end{tabular}
\end{ruledtabular}
\end{table}

\begin{center}
{\bf {IV}. DISCUSSION AND CONCLUSION}
\end{center}

The ionic radius of Ni$^{2+}$ (0.70 \AA) is smaller than that of Mn$^{2+}$ (0.80 \AA), so the effect of increasing
Ni$^{2+}$ content on the nuclear structure is to be expected. As exhibited 
in Table 1, $a$ and $c$ both decrease with 
increasing Ni-doping, so do the $z$ values of the atoms on $6c$ sites (Mn, Ni, and Ti). 
The value of $O_y$ for the oxygen site, already small in MnTiO$_3$ (0.031), is systematically reduced 
by the Ni-replacement and becomes 0.016 in NiTiO$_3$.
However, its minuscule value keeps the crystal from having mirror planes, so 
is important for the crystallographic symmetry and consequently
for the magnetic symmetry. 
The effective moment for Mn$^{2+}$ in MnTiO$_3$, 4.55 $\mu_B$, is smaller than the spin-only value.  
This can be ascribed to the incomplete ordering of Mn and Ti or the existence of Mn$^{3+}$. \cite{Shirane}  Both the effective moment 
and the N$\acute{e}$el temperature change with Ni concentration as a result of the competing anisotropies and frustrated exchange interactions . 
This change is more rapid on the Mn-rich region. Both values are considerably reduced at x=0.5, which is compatible with the observed 
spin glass behavior. In the x=0.68 compound, the intra-layer exchange interactions among the Ni$^{2+}$ ions become so dominant that 
$T_N$ and the ordered moment are very close to those in the pure NiTiO$_3$. Because of the similar radii of Ni and Ti ions, more 
incomplete ordering exists in NiTiO$_3$ \cite{Ishikawa}, which is mainly responsible for the less-than-expected moment size of Ni$^{2+}$. \cite{Shirane} 

The Mn$^{2+}$ and Ni$^{2+}$ ions have distinct single ion anisotropies 
as manifested by their different easy axes in the ilmenites \cite{Ishikawa,Shirane} and 
other compounds such as barium fluorides BaMnF$_4$ \cite{Poole} and BaNiF$_4$. \cite{Cox}
The added Ni cations randomly replace Mn on the octahedral sites 
and weaken the spin correlations, more so in the inter-planar direction, as indicated 
by the enhanced diffuse scattering in the x=0.33 system.
Although the spin correlation starts to form high above $T_N$,
the electric polarization does not occur until the long range G-type magnetic order 
is established. 
When the spins are parallel to the $c$-axis, the magnetic group symmetry is
$R\bar{3}'$ and the point symmetry is  $\bar{3}'$.  
As the collinear AFM moments tilt away from the $c$ axis, the emerged $a$-axis components 
in the hexagonal layer
loses the 3-fold rotation symmetry. The magnetic space group then becomes $P\bar{1}$ 
and the magnetic point symmetry becomes $\bar{1}'$. 
Since the nonzero $O_y$ value 
keeps the crystal from having 2-fold rotation axis and mirror planes. Even if
the spins completely lie in the $a$-axis, the point group of the magnetic 
symmetry is not $2'/m$ as it appears to be. The restrictions
from the non-magnetic anion sites must be obeyed as the Neumann's principle
requires the physical property tensor 
be invariant under all the permissible operations of the crystallographic symmetry \cite{Birss, Bhagavantam}. 

This observation is the key to understand the observed electric polarizations 
summarized in Fig.2 and Fig.3 for x=0 and 0.33, respectively.
Both $\bar{3}'$ and $\bar{1}'$ are among the 
58 magnetic point groups that have non-zero elements in their
ME susceptibility tensors. \cite{Rivera}
The former has both diagonal $\alpha_{xx}$=$\alpha_{yy}$, $\alpha_{zz}$ and off-diagonal
components $\alpha_{xy}$=-$\alpha_{yx}$, while the latter does not impose any restrictions on 
the form of ME tensor and all tensor components are non-zero. 
The G-type structure with $c$-axis spins ($\phi$=0) in MnTiO$_3$,
permits
$\alpha_{zz}$ as indeed observed.
In the x=0.33 system, the spins tilt away from $c$ while maintaining the $G$-type structure
and lowers the symmetry to $\bar{1}'$. The symmetry remains as $\bar{1}'$ even for $T_R$$<$$T$$<$$T_N$ 
unless a $c$-direction magnetic field pull the spins back along $c$ (Fig. 6(d)), which enables the recovery of 
the $\bar{3}'$ symmetry in this temperature range. This explains why $\alpha_{zz}$, allowed by both 
symmetries, exist in the entire $T$$<$$T_N$ range. 
Cooling across $T_R$ at 20 K, $\bar{1}'$ arising from
the collinear spin rotation triggers $\alpha_{xz}$, which is prohibited by $\bar{3}'$.  
At the same temperature, 
$\alpha_{zz}$ exhibits considerable suppression due to the reduced
$c$-component of the moment, as shown by the red triangle in Fig.6(d). 
The change of  
magnetic point symmetry satisfactorily 
explains the temperature dependence of the observed $P_a$ and $P_c$. It is clear that  
coupling of the ferroelectric order and magnetic order is due to the linear ME effect.
The case of x=0.33 system is different from a normal linear ME effect, as in x=0, in that
the spin directions vary with external magnetic field, which fails the linear dependence 
of the polarization on magnetic field.  

With the ties between the two orders established, one can use the polarization to 
predict the spin structures at higher fields, as they are difficult to determine
experimentally. The representation
analysis using SARAh program \cite{Wills} shows that  
for the space group $R\bar{3}$ with magnetic propagation vector \textbf{\textit{k}}=(0,0,0), the G-type 
is the only possible AFM spin arrangement. So if one assumes the magnetic
wave vector remains unchanged, $\phi$ alone should
be sufficient to describe all the spin structures under moderate magnetic field.
With higher $H_{\|c}$ in the x=0.33 compound, $\phi$ remains different in the two temperature regions. 
The fact that $P_a$ only exists below $T_R$ (Fig.2(d)) suggests that
up to $H_{\|c}$=7 T, $\phi$=0 for $T_R$$<$T$<$$T_N$ and that $\phi\neq$0 for $T$$<$$T_R$.
As the magnetic field is applied along $a$,  the observed $\alpha_{xz}$ (Fig.3(d)), 
prohibited by $\bar{3}'$, suggests the $\bar{1}'$ magnetic point symmetry.

The electric polarization flop has been 
observed in a few multiferroic materials, including rare-earth manganites  RMnO$_3$, \cite{Kimura2, Strempfer} 
RMn$_2$O$_5$ \cite{Fukunaga} and the mineral h$\ddot{u}$bnerite MnWO$_4$, \cite{Taniguchi}
which generally have incommensurate noncollinear spiral spin structures. In these materials the $P$-flop
is typically caused by the flop of spiral or cycloid plane. 
MnTiO$_3$ is a rare case of magnetic field induced $P$-flop with a collinear magnetic structure.  
In the x=0.33 system,
the polarizations in the two directions are turned on by the same field and coexist for $T$$<$$T_R$,
so this is not a typical $P$-flop. But the reciprocal interactions between $P_a$ and $P_c$ and their 
different on-set temperatures makes it a unique type of ME control.
The Co-doped MnWO$_4$ is another case of  
$P$-flop caused by the competing single ion anisotropies, which is achieved by the flop of the spin helix. \cite{Liang,Ye} 
But the magnetic frustration and complex magnetic structure
make this type of control difficult to repeat in other compounds in terms of materials design.  
In comparison, the collinear spin rotation
in Mn$_{1-x}$Ni$_x$TiO$_3$ can be easily created for 
a random mixture of two antiferromagnets with orthogonal easy axes.  
A new intermediate phase whose easy axis tilts
oblique to the easy axes of the pure systems, and two second order transitions are all predicted 
by mean field approximation. \cite{Matsubara, Oguchi}
Such predictions have also been fulfilled in other random mixtures such as K$_2$Mn$_{1-x}$Fe$_x$F$_4$ \cite{Bevaart} 
and Co$_{1-x}$Fe$_x$Cl$_2$2H$_2$O. \cite{Kobayashi}

\begin{center}
{\bf V. ~~SUMMARY}
\end{center}

The structural, magnetic and electric properties have been
studied for 4 typical compositions of Mn$_{1-x}$Ni$_x$TiO$_3$. 
Magnetic field induced electric polarizations have been observed in the compositions x=0 and 0.33, both
of which have the G-type magnetic order. In the x=0 system, the polarizatio flops from $P_c$
to $P_a$ as the spin flop transition is triggered at $H_{\|c}$=7 T.
In x=0.33, $P_a$ is turned on together with $P_c$ by $H_{\|c}$. Additionally, $P_a$ and $P_c$ 
can also be induced by $H_{\|a}$. 
By studying the magnetic structure and phase transition with and without magnetic field, 
the occurrence of the new ME coupling is attributed to the emergent point group symmetry as the antiferromagnetically 
coupled spins tilt collinearly toward the $a$-axis. Such spin rotation results from the strong competition of 
single ion anisotropy of the transition metal elements and provides a new way to tune 
electric polarizations.
The magnetic structure of the x=0.5 and 0.68 systems is the same as that of the NiTiO$_3$. No polarization was observed.   

\begin{center}
{\bf V. ~~ACKNOWLEDGMENTS}
\end{center}

The research at Oak Ridge National Laboratory's High Flux Isotope Reactor was sponsored by the
Scientific User Facilities, Office of Basic Energy Sciences, US Department of
Energy. The authors are grateful for fruitful discussions with Bryan C. Chakoumakos.
H.D.Z thanks for the support from JDRD program of University of Tennessee. 
NHMFL is supported by National Science Foundation (DMR-0654118) , the State of Florida, and the U.S. Department of Energy.

\end{document}